\documentclass[12pt]{iopart}
\usepackage{widetext}
\usepackage{eqnarray}
\usepackage{iopams}  
\usepackage{graphicx, caption}
\usepackage{setspace}
\bibstyle{iopart-num}
\usepackage{etoolbox}
\newtoggle{SUPPORT}
\togglefalse{SUPPORT}
\long\def\support#1{ \iftoggle{SUPPORT}{\begin{widetext}\small {\leavevmode\color{red}{#1}} \normalsize\end{widetext}}{} }
\begin{document}
\maketitle
\title[Exceeding the Sauter-Schwinger limit of pair production with a quantum gas]{Exceeding the Sauter-Schwinger limit of pair production with a quantum gas}

\author{A.~M.~Pi\~{n}eiro, D.~Genkina, Mingwu~Lu and I.~B.~Spielman}
\address{Joint Quantum Institute, National Institute of Standards and Technology and University of Maryland, Gaithersburg, MD 20899, USA}
\ead{\mailto{pineiro@.umd.edu}, \mailto{ian.spielman@nist.gov}}

\begin{abstract}
We quantum-simulated particle-antiparticle pair production with a bosonic quantum gas in an optical lattice by emulating the requisite 1d Dirac equation and uniform electric field. We emulated field strengths far in excess of Sauter-Schwinger's limit for pair production in quantum electrodynamics, and therefore readily produced  particles from ``the Dirac vacuum'' in quantitative agreement with theory. The observed process is equivalently described by Landau-Zener tunneling familiar in the atomic physics context.
\end{abstract}

%
%
%


The creation of particle-antiparticle pairs from vacuum by a large electric field is a phenomenon arising in quantum electrodynamics (QED)~\cite{Dirac1928,Sauter1931,Heisenberg1934,Heisenberg1936}, with threshold electric field strength ${E_{\rm c} \approx 10^{18}\ {\rm V}/ {\rm m}}$ first computed by J. Schwinger~\cite{PhysRev.82.664}. Electric fields on this scale are not experimentally accessible; even the largest laboratory fields produced by ultrashort laser pulses~\cite{Bulanov2010} fall short, making direct observation of pair production out of reach of current experiments. To experimentally probe this limit with a bosonic quantum gas, we engineered the relativistic 1d Dirac Hamiltonian with $mc^{2}$ reduced by $17$ orders of magnitude, allowing laboratory scale forces to greatly exceed Sauter-Schwinger's limit. We readily measured pair production and demonstrated that this high-energy phenomenon is equivalently described by Landau-Zener tunneling~\cite{Landau, Zener696}.

In the Dirac vacuum, the enormous electric field required is 
\begin{eqnarray}
{E_{\rm c} = m_{\rm e}^2c^3/\hbar q_{\rm e}} ,
\end{eqnarray}
determined by the particle/antiparticle mass $m_{\rm e}$ and charge $q_{\rm e}$. For an applied electric field $E$, the pair production rate is governed only by the dimensionless ratio ${E/E_{\rm c}}$, allowing our physical system with very different characteristic scales to be used to realize the underlying phenomenon.  

Our system was well described by the 1d Dirac Hamiltonian~\cite{Gerritsma_2010,Weitz_2010,Weitz_2011,LeBlanc_2013}
\begin{eqnarray}
\hat{H}_{\rm D} = c\hat{p} \sigma_z +  m c^2 \sigma_x,
\end{eqnarray}
where $\hat{p}$ is the momentum operator and ${\sigma_{x,y,z}}$ are the Pauli operators. Starting with $m=0$, the 1d Dirac Hamiltonian describes particles with velocities $\pm c$, that are then coupled with strength $mc^2$ to give the familiar ${\mathcal{E} (p) = \pm (p^{2} c^{2} + m^{2} c^{4})^{1/2}}$ dispersion relation for relativistic particles and antiparticles. At zero momentum, this dispersion has a gap equal to twice the rest mass, at which point the curvature is inversely proportional to the rest mass. The Dirac vacuum consists of occupied states in the lower (antiparticle) band of the Dirac dispersion and vacant states in the upper (particle) band. Vacancies in the antiparticle band represent antiparticles and occupied states in the particle band represent particles. We probed Schwinger's limit for pair production by measuring the probability and rate of `pairs' out of this vacuum as a function of rest mass and applied force. 

We emulated $\hat{H}_{\rm D}$ with the lowest two bands of a 1d optical lattice, giving the pair of relativistic modes at the edge of the Brillouin zone shown in figure 1. The rest mass ${m^{*}c^{*2} = V/4}$ is set by the peak-to-valley lattice depth $V$ generated by our ${\lambda_{\rm L} = 1064\  {\rm nm}}$ laser light. The speed of light is replaced by the greatly reduced speed of light ${{c^{*} = \hbar k_{\rm{L}}/m_{\rm Rb} \approx 4.3\  {\rm mm/s}}}$ equal to the single photon recoil velocity. The single photon recoil momentum ${\hbar k_{\rm{L}} = 2\pi \hbar/\lambda_{\rm{L}}}$ specifies the recoil energy ${E_{\rm{L}} = \hbar^2 k_{\rm{L}}^{2}/2m_{\rm Rb} = h \times 2.02\  {\rm kHz}}$. These recoil units set the scale for all physical quantities in our analog system. The effective Compton wavelength ${\lambda_{\rm{C}} = h/m^{*}c^{*} = 8 \lambda_{\rm L} E_{\rm L}/V}$ is about $10^{6}$ times larger than that of an electron.


\begin{figure}
\includegraphics[width=\columnwidth,clip=]{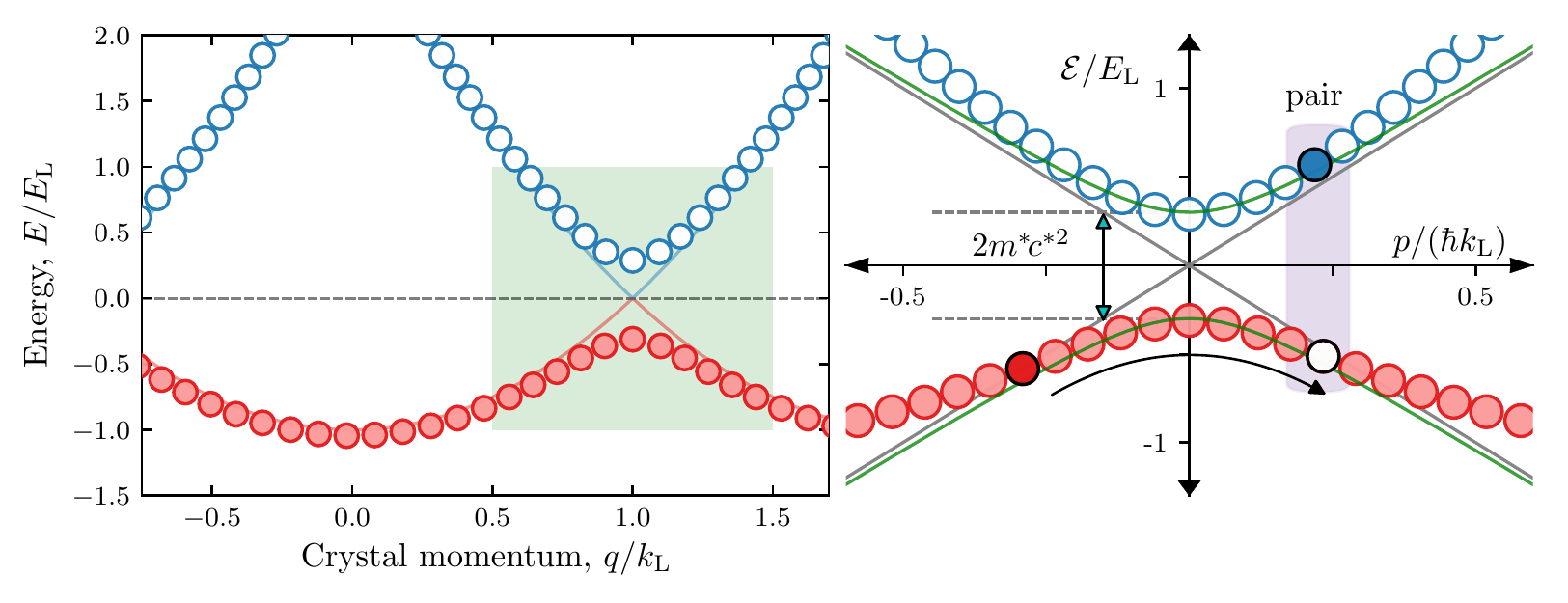}
\caption{1d Dirac equation derived from optical lattice band structure. The left panel shows the lowest two bands of a 1d optical lattice and right panel shows the expanded view at the edge of the Brillouin zone. In both cases, the dispersion of the lower (or ``antiparticle'') band is denoted by red circles and of the upper (or ``particle'') band is denoted by blue circles. The expanded view reveals the connection to the 1d Dirac equation (green shaded region in left panel).  The gray lines plot the linear dispersion of massless particles, while the green curves depict the Dirac-dispersion computed from (2). The curved arrow illustrates the process of pair production where an occupied state in the antiparticle band is converted to an occupied state in the particle band and a vacancy in the antiparticle band.}
\end{figure}   

Our simulations consisted of ultracold bosons first prepared in the antiparticle band, then subjected to a constant force, ${F_{\rm e} = q_{\rm e} E = \hbar dq/dt}$, modeling an electric field. During the application of this force atoms may transfer from the antiparticle band to the particle band, emulating the pair production phenomenon. We measured the fraction of atoms transferred to the particle band as a function of the effective rest mass and the applied force. This transfer between bands is described by Landau-Zener tunneling. In this manuscript, we begin by using a Bose-Einstein condensate (BEC) to illucidate the connection between Landau-Zener tunneling and pair production; we then model the Dirac vacuum by uniformily filling the Brillouin Zone of the antiparticle band and observe the predicted rate of pair production.


Our experiments began with nearly pure $^{87}$Rb Bose-Einstein condensates (BECs) in the ${\vert F = 1, m_{F} = -1 \big>}$ internal state, in a crossed optical dipole trap~\cite{Lin2009} formed at the intersection of two laser beams traveling along ${\bf e}_{x}$ and ${\bf e}_{y}$, giving trap frequencies ${(f_x, f_y, f_z) = (44, 45, 94) \ {\rm Hz}}$. The low density of our ${N\! \approx10^3}$ atom BECs limited unwanted scattering processes in regimes of dynamical instability~\cite{PhysRevLett.96.020406}. The optical lattice potential was formed by a retro-reflected ${\lambda = 1064\ {\rm nm}}$ laser beam with a waist of ${\approx 150\ \mu m}$. Emulated electric forces were applied by spatially displacing the optical dipole beam providing longitudinal confinement (by frequency shifting an acousto-optic modulator). This effectively added a linear contribution to the existent harmonic potential for displacements small compared to the beam waist.

We loaded BECs into the optical lattice by linearly increasing the lattice laser intensity from zero to the final intensity in $300\ {\rm ms}$, a time-scale adiabatic with respect to all energy scales.  
Once the final lattice depth---determined by the laser intensity$\!$$\!^{~\footnote{The lattice was calibrated by using Kapitza-Dirac diffraction of the BEC off a pulsed lattice potential.~\cite{Kapitza-Dirac-diffraction}}}$---was achieved, we applied a force for a time $t_{F}$. Immediately thereafter, the lattice was linearly ramped off in $1\ {\rm ms}$, mapping crystal momentum states to free particles states~\cite{PhysRevLett.74.1542,PhysRevLett.87.160405,PhysRevLett.94.080403,RevModPhys.80.885}. This process mapped atoms in the anti-particle band to free particle states with momentum between ${-k_{\rm L}}$ and ${k_{\rm L}}$, and mapped atoms in the particle band to states ${\pm k_{\rm L}}$ and ${\pm 2k_{\rm L}}$. The resulting momentum distribution was absorption imaged after a $15.7\ {\rm ms}$ time-of-flight (TOF).


\begin{figure}[t]
\centering
\includegraphics[height=.8\columnwidth,clip=]{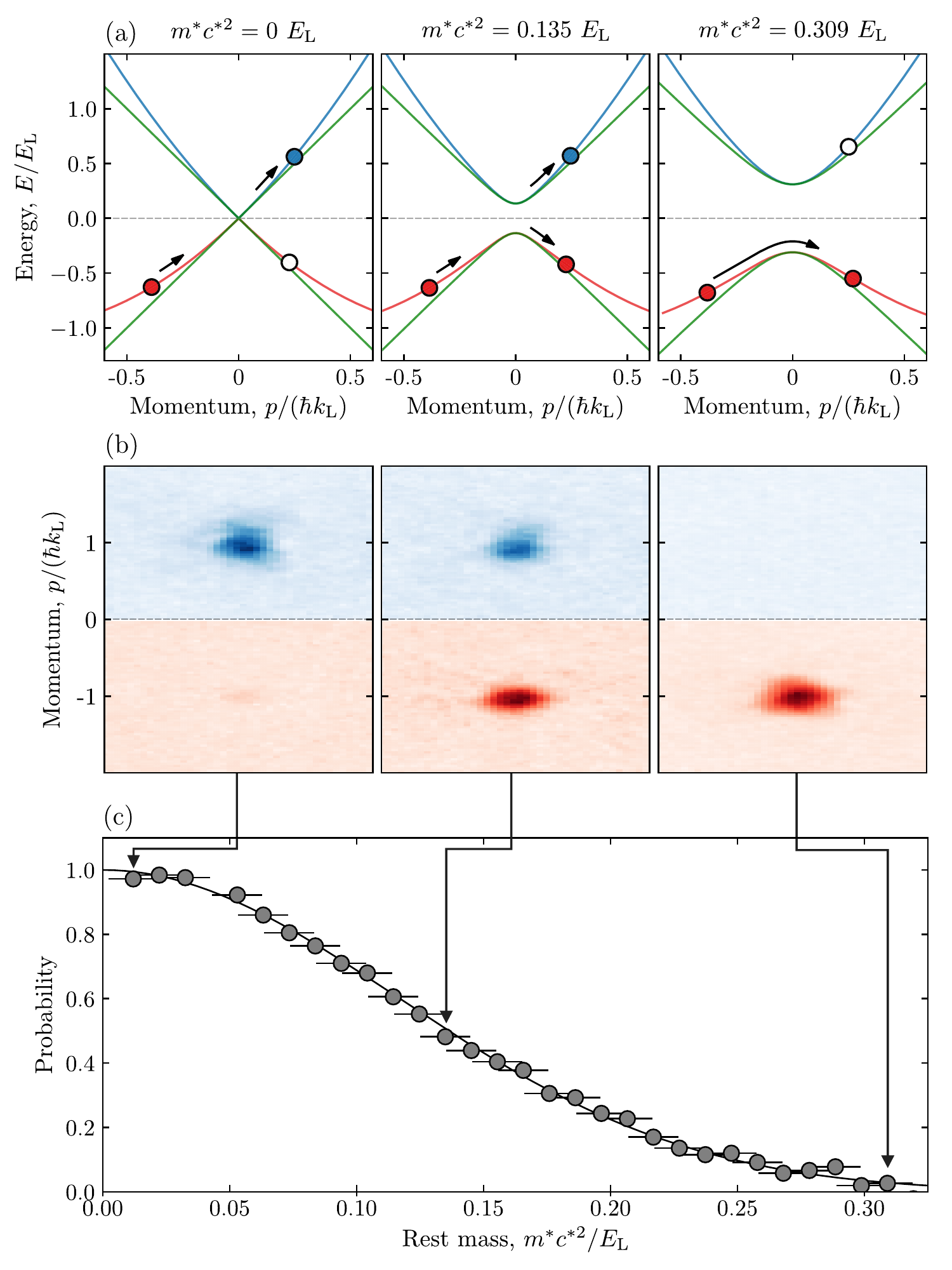}
\captionsetup{width=.9\linewidth} 
\caption[fig2]{Pair production from a single state with electric force ${F_{\rm e} = 2\hbar k_{\rm L}/t_{F}}$ applied for a time ${t_{F} = 3.7\ {\rm ms}}$. (a) Schematic representation of pair-production for rest mass = $(0.012(1), 0.134(9), 0.309(1)) E_{\rm L}$ respectively.  In each case the green curves denote the Dirac dispersion, and the blue and red curves denote the particle and anti-particle bands respectively. (b) TOF images showing the fractional populations of atoms occupying the particle band (blue tone) and antiparticle band (red tone) for each $m^*$. (c) Probability of pair production as a function of rest mass, plotted along with the Landau-Zener model from (3). Statistical uncertainties are shown with typical error bars.}
\end{figure} 

We began with an experiment that is natural in the cold atom setting but is unphysical in high energy physics: we applied an effective electric field and varied the rest mass. The result of this changing mass is schematically shown in figure 2(a) by an increasing gap at zero momentum. The decreasing probability of pair production with increasing mass anticipated by (1) is schematically illustrated by the filled or partially filled circles. Figure 2(b) shows the distribution of atoms at time $t_{F}$ selected so that the atoms traversed a full Brillouin zone, i.e., underwent a single Bloch oscillation~\cite{PhysRevLett.76.4508}. Atoms in the top portion of the panel (blue tone) represent particles and atoms in the bottom portion of the panel (red tone) denote filled vacuum states. Figure 2(c) quantifies this effect in terms of the fractional population of a BEC transferred into the particle band, and as expected, the probability of pair production monotonically decreased with increasing effective rest mass.  For small rest masses, the atoms almost completely populated the particle band, while for large mass the particle band was nearly empty.

\begin{figure}
\includegraphics[width=\columnwidth,clip=]{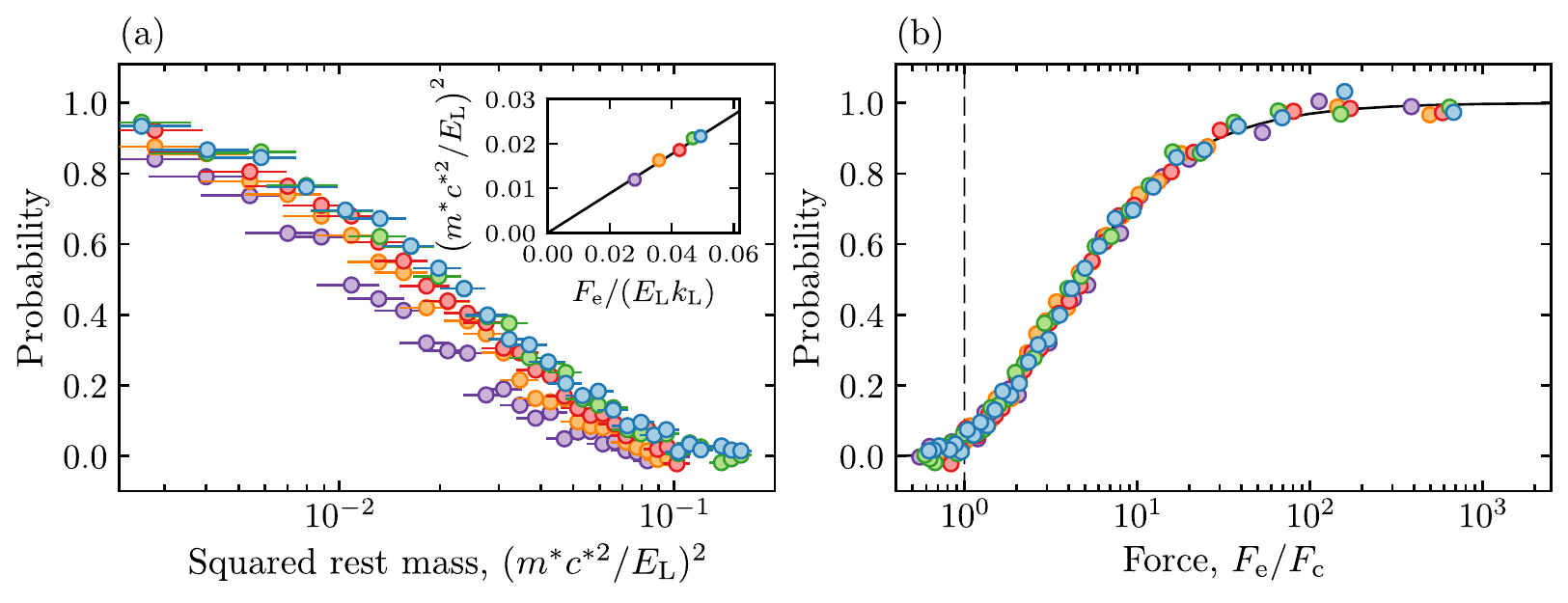}
\caption[fig3]{(a) Probability of pair production as a function of rest mass squared for ${F_{\rm e} = 2\hbar k_{\rm L}/t_{F}}$ and ${t_{F} = (3.2, 3.4, 3.7, 4.4, 5.6)\ {\rm ms}}$, showing the same functional form. Statistical uncertainties are shown with typical error bars. The inset quantifies the scaling relationship between these curves in terms of the point where each curve reaches half-max, in agreement with the prediction of (3), shown by the black line. (b) Probability of pair production plotted as a function of dimensionless force ${F_{\rm e}/F_{\rm c}}$, showing collapse onto a single curve.  The vertical dashed line marks Schwinger's limit where ${F_{\rm e} = F_{\rm c}}$, and the solid curve is the prediction of (3).}
\end{figure} 

The solid curve in figure 2(c) plots the Landau-Zener diabatic transition probability~\cite{Landau, Zener696, PhysRevA.23.3107, Raizen_1998, Weitz_2010} given by
\begin{eqnarray}
P_{\rm LZ} = e^{-2\pi\Gamma}, \textrm{with}\ \Gamma = \frac{a^{2}}{\hbar |\frac{d}{d t}\Delta E|}, 
\end{eqnarray}
\support{Taylor expanding about $q=k_{\rm L}$ for the two lowest bare eigen-energies of the 1d lattice Hamiltonian gives ${E(\rm q) = E_{\rm L} \pm 2E_{\rm L}(q - 1)}$.
\\ Landau-Zener expression:
\begin{eqnarray}
P_{LZ} = e^{-2\pi\Gamma}, with\ \Gamma = \frac{a^2}{\hbar |\frac{d}{d t} E_{2}-E_{1}|} 
\end{eqnarray}
\\ Modified Landau-Zener expression:
\begin{eqnarray}
\Gamma = \frac{V^2}{16\hbar |\frac{d}{d t} [4 E_{\rm L}(q -1)]|} 
= \frac{V^2}{64\hbar |E_{\rm L} \frac{d q}{d t}|}  
\end{eqnarray}}
describing the transit through a crossing with gap $a=2 m^*c^{*2}$ while the massless energy difference $E(p) = 2 c^* p$ changes as $p$ is swept at constant rate $dp/dt$. The rate ${\hbar dq/dt}$ sets the electric force ${F_{\rm e} = q_{\rm e} E}$. Remarkably in terms of these parameters, the Landau-Zener coefficient is defined by ${F_{\rm e}/F_{\rm c} = 1/2 \Gamma}$ where ${F_{\rm c} = q_{\rm e} E_{\rm c}}$ is exactly Schwinger's limit of pair production. This prediction is in near perfect agreement with data in figure 2(c).


As suggested by the quadratic dependance on mass in (1), figure 3(a) plots the probability of pair production as a function of rest mass squared for 5 different forces, illustrating their similar behavior. The inset figure confirms the agreement with the Landau-Zener expression by ploting the rest mass required to achieve a ${50 \%}$ probability of pair production as a function of the field strength, and the solid curve shows good agreement with the Landau-Zener prediction. Figure 3(b) displays the same data (circles) now as a function of ${F_{\rm e}/F_{\rm c}}$ that collapses onto the predicted transition probability (solid curve). This collapse confirms that the physics of pair production exhibits universal behavior when the field is expressed in units of $F_{\rm c}$ as predicted by the Landau-Zener expression. The vertical dashed line marks the ratio ${F/F_{\rm c} = 1}$, Sauter-Schwinger's limit for pair production; much of our data is in the high-field limit, which is extremely difficult to achieve in other physical contexts. Our data densly samples the critical threshold regime near $F_{\rm c}$ and spans the full gamut from vanishing pair prduction to nearly complete pair production. Together these data show the clear connection between the Landau-Zener tunneling of a single quantum state and the phenomenon of pair production. Still, the single occupied state defined by our BEC is far from the Dirac vacuum state. 

\begin{figure}
\includegraphics[width=\columnwidth,clip=]{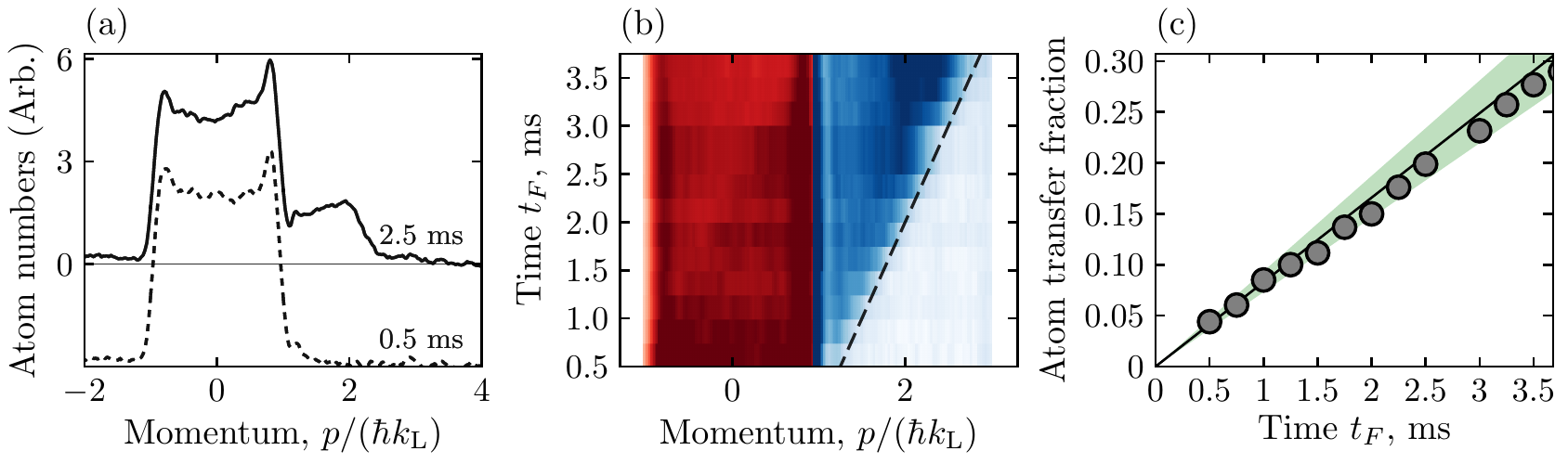}
\caption{Time dependence of pair production from the Dirac vacuum with ${F_{\rm e}/F_{\rm c} = (2.3(6), 2.7(5), 3.3(1), 4.1(3), 5.5(1), 8.2(6), 16.5(3))}$.$\!$ (a)$\!$ Momentum distribution following the application of a force for short (dashed) and long (solid) hold times.$\!$ (b) Time-dependence of pair-production showing the appearance of and acceleration of particle states and the depletion and deceleration of antiparticle states as a function of time. The red regions and blue regions mark states in the anti-particle and particle bands respectively. The dashed line marks the acceleration expected from the applied force alone. Both the slight wedge shape of the particle-state distribution in (a) and the slight shift from the dashed line in (b) result from the harmonic confining potential.$\!$ (c) Observed fractional transfer of atoms as a function of time plotted along with the prediction of (3).$\!$ The green shaded area represents the uncertainty from input parameters.}
\end{figure} %
We therefore created an initial state mimincking the Dirac vacuum in which the negative energy states were occupied with equal probability and the positive energy states were vacant. We prepared this state by first adiabatically loading a BEC into a fairly deep optical lattice (${V \approx 3.4E_{\rm L}}$, making the lowest bands well separated) and applied a force for $1\ {\rm s}$, sufficient for about 300 Bloch oscillations to take place~\cite{PhysRevLett.76.4508}. During this time, crystal momentum changing collisions~\cite{PhysRevLett.96.020406} and dephasing processes uniformly filled the antiparticle band. We then adiabatically reduced the lattice depth in $300\ {\rm \mu s}$ giving $m^{*}c^{*2}/E_{\rm L} = 0.2(1)$, and proceeded as described in the experimental methods section. Here, pairs were produced at a constant rate for the entire time $t_{F}$ the force was applied, filling initially vancant states in the particle band while depleting initially occupied states in the antiparticle band. For these data ${F_{\rm e}/F_{\rm c} = (2.3(6), 2.7(5), 3.3(1), 4.1(3), 5.5(1), 8.2(6), 16.5(3))}$. Figure 4(a) shows the atomic momentum distribution following this procedure for small (dashed curve) and larger (solid curve) values of $t_F$.  In these data the states in the antiparticle band fall in the pink region, and states in the particle band fall in the blue regions. A uniformly filled initial band would ideally produce a top-hat distribution in the pink region; we confirmed that the observed time-independent peaks in the distribution are consistent with a slight defocus of our imaging system~\cite{Turner_thesis, Turner_2004, Putra_2014}.  The evolution from the dashed curve to the solid curve clearly shows particles that have been created and then accelerated by $F_{\rm e}$. Figure 4(b) summarizes data of this type for variable $t_F$, making clear the appearance of particles already visible in figure 4(a), but also showing a similar reduction of occupation in the antiparticle band. The white dashed line marks the anticipated acceleration given by $F_{\rm e}$; the slightly reduced observed acceleration results from the harmonic confining potential. Finally, figure 4(c) plots the fractional occupation probability of the particle band, clearly showing the production of pairs at a constant rate.  The line plots the rate, ${dq/dt = 2\hbar k_{\rm L}/t_{F}}$ and ${t_{F} = 3.4\ {\rm ms}}$, derived from the transition probability in (3).  This then is the direct analog of the constant rate of pair production from vacuum from a uniform electric field. 


In this experiment, we used a cold atom system to quantitatively probe both the underlying mechanism and the overall phenomena of pair production as initially conceived of in high-energy physics.  Our analysis shows that pair production can be equivalently understood as a quantum tunneling process, and our data spans the full parameter regime from low applied field (negligible pair production) below the Sauter-Schwinger limit, to high field (maximum rate of pair production) far in excess of the Sauter-Schwinger limit.  High-intensity pulsed laser experiments~\cite{PhysRevLett.101.200403, PhysRevLett.102.080402, Kirk_2009, Bulanov2010, Hill_2017} promise to measure the vacuum non-linearity and ultimately exceed Sauter-Schwinger's limit in its original context.  Current theory suggests that the actual threshold for pair production will be somewhat in excess of $E_{\rm c}$ resulting from the Coulomb attraction between electron-position pairs.  Future cold atom experiments with repulsively interacting fermions could probe this ``excitionic'' shift as well, allowing more quantitative comparison with higher order corrections to the threshold field strength. In addition, the pair-production phenomena occurring in strongly interacting field theories, even absent applied electric fields, may also be realized using mixtures of ultracold bosons and fermions~\cite{Kasper2016, Kasper2017}.

This work was partially supported by the National Institute of Standards and Technology, Air Force Office of Scientific Research's Quantum Matter Multidisciplinary University Research Initiative, and the National Science Foundation through the Physics Frontier Center at the Joint Quantum Institute. A.M.P. was supported by the U.S. National Science Foundation Graduate Research Fellowship Program under Grant No. DGE 1322106.\\

\section*{References}
\bibstyle{iopart-num}
\bibliography{SPP}

\end{document}